\documentclass[review,12pt]{elsarticle}
\usepackage{amsmath, amsthm, amssymb}

\begin{document}
\title{Improving procedure for the reconstruction of an effective secular equation}
\author{Yong Zheng  \corref{correspondingauthor}}
\cortext[correspondingauthor]{Corresponding author}
\ead{zhengyongsc@sina.com}
\address{School of Physics and Electronics, Qiannan Normal University for Nationalities,
	Duyun 558000, China}

\begin{abstract}
The usual implement procedure for the reconstruction of secular equation for an effective Hamiltonian has been discussed and improved. A relative characteristic polynomial has been introduced for the effective Hamiltonian, to  obtain a simplified effective-secular-equation which is equivalent to the one obtained in the usual reconstruction  procedure but shows great convenience and effectiveness, especially when the $P$-space dimension is large.
\end{abstract}

\begin{keyword}
Effective secular equation \sep Effective Hamiltonian \sep Resummation method
\sep Improving procedure
\end{keyword}

\maketitle

\section{Introduction}

Recently, we have studied the problem how to enlarge the convergence radius in a perturbation treatment of an effective Hamiltonian  $ H_{\mathrm{eff}}(\lambda) $, with $ \lambda $ the perturbation parameter \cite{Z}. Our treatment is based on the relatively good analytic property of the characteristic polynomial of the effective Hamiltonian $\det[E-H_{\mathrm{eff}}(\lambda)]$. Fern\'{a}ndez has reminded that similar treatment had been developed several years earlier by Fried and Ezra, via the reconstruction of an effective secular equation for $ H_{\mathrm{eff}}(\lambda) $  \cite{C,Fried}. 
The differences between these two treatments, the one of Fried and Ezra and ours,  have been analyzed in our previous reply to Fern\'{a}ndez's comment \cite{Z2}, especially the  complex-energy-eigenvalue problem associated  with Fried and Ezra's treatment.  Here, we want to give a further discussion on how to get a more effective solution procedure based on Fried and Ezra's treatment, to reconstruct an effective secular equation for $ H_{\mathrm{eff}}(\lambda) $. 

\section{Usual implement of the reconstruction of an effective secular equation}
For the convenience of discussion, we first give a quick look at Fried and Ezra's treatment for the reconstruction of an effective secular equation \cite{C,Fried}. 

As in \cite{Z}, We can also discuss in a state space with a total dimension $ \tilde{N}=N+M $, where $N$ is the dimension of the so-called $P$-space spanned by states we want to study,  and $M$ is that of the $Q$-space spanned by the other states. A general form of the Hamiltonian can be written as
\begin{equation}\label{H}
	H(\lambda)=H_0+\lambda H_I,
\end{equation}
with
\begin{align*}
	&H_0=\sum_{n}\epsilon_{n}^{_P}|\psi^{_P}_n\rangle \langle \psi^{_P}_n|+\sum_{m} \epsilon_{m}^{_Q}|\psi^{_Q}_m\rangle \langle \psi^{_Q}_m|,\\
	&H_I=\sum_{n,n'} h^{\!_P}_{nn'}|\psi^{_P}_n\rangle \langle \psi^{_P}_{n'}|+\sum_{m,m'} h^{\!_Q}_{mm'}|\psi^{_Q}_m\rangle \langle \psi^{_Q}_{m'}|  +\sum_{n,m}  \left[ h^{\!_{P\!Q}}_{nm}|\psi^{_P}_n\rangle \langle \psi^{_Q}_m|+\text{h.c}\right],
\end{align*}
where the indexes in the summations: $n(n') = 1,2,\cdots, N$ and $m(m')= 1,2,\cdots, M$. 
We also assume here that the unperturbed energies all are non-degenerate (though this is not necessary, as has been discussed in \cite{Z}) for the convenience of later discussion, i.e.,  $ \epsilon_1^{_P} \neq \epsilon_2^{_P}$, $ \epsilon_1^{_P} \neq \epsilon_1^{_Q}$, etc.

To obtain the eigenvalues of $P$-space states, $ E_1^{_P}(\lambda ), E_2^{_P}(\lambda ),\cdots, E_{N}^{_P}(\lambda )$, one can construct the so-called effective secular equation:
\begin{equation}\label{edetN}
	\det[E-H_{\mathrm{eff}}(\lambda)]=\prod_{n=1}^{N}[E- E_n^{_P}(\lambda )]=0,
\end{equation}
Obviously, the specific form of this equation is unknown until the exact $P$-space eigenvalues all have been obtained. Hence, a reconstruction procedure is employed. One first calculates the $P$-space eigenvalues perturbatively,
\begin{subequations}
	\begin{align}
		&E_1^{_P}=\epsilon_1^{_P}+\lambda h^{\!_P}_{11}+ \lambda^2\left( \sum_{n\neq
			1}\frac{|h^{\!_P}_{1n}|^2}{\epsilon_1^{_P}-\epsilon_n^{_P}}
		+ \sum_{m}\frac{|h^{\!_{P\!Q}}_{1m}|^2}{\epsilon_1^{_P}-\epsilon_m^{_Q}}\right) +\cdots \tag{3.1} \label{EP1}\\
		&E_2^{_P}=\epsilon_2^{_P}+\lambda h^{\!_P}_{22} + \lambda^2\left( \sum_{n\neq 2}\frac{|h^{\!_P}_{2n}|^2}{\epsilon_2^{_P}-\epsilon_n^{_P}}
		+ \sum_{m}\frac{|h^{\!_{P\!Q}}_{2m}|^2}{\epsilon_2^{_P}-\epsilon_m^{_Q}}\right) +\cdots  \tag{3.2} \label{EP2}\\
		&\quad \cdots \cdots \notag\\
		&E_N^{_P}=\epsilon_N^{_P}+\lambda h^{\!_P}_{N\!N}+ \lambda^2\left(
		\sum_{n\neq
			N}\frac{|h^{\!_P}_{Nn}|^2}{\epsilon_N^{_P}-\epsilon_n^{_P}} +
		\sum_{m}\frac{|h^{\!_{P\!Q}}_{Nm}|^2}{\epsilon_N^{_P}-\epsilon_m^{_Q}}\right)
		+\cdots.  \tag{3.N} \label{EPN}
	\end{align}
\end{subequations}
Then, one can substitute these $ \lambda $-series form of $ E_n^{_P} $ (no worries about the convergence radius, according to the discussion in \cite{Z}) into Eq.~\eqref{edetN} to reconstruct the effective secular equation to needed order of $ \lambda $, say, 
\begin{align*}
	&\text{to  $ \lambda^0$: } \{\det[E-H_{\mathrm{eff}}(\lambda)]\}^{[0]}= \prod_{n}[E- \epsilon_n^{_P}]=0; \\
	&\text{to  $ \lambda^1$ }: \{\det[E-H_{\mathrm{eff}}(\lambda)]\}^{[1]}= \prod_{n}[E- \epsilon_n^{_P}]-\lambda \sum_{n}h^{\!_P}_{nn}\prod_{n'\neq n}[E- \epsilon_{n'}^{_P}]=0; \\
	&\text{to  $ \lambda^2$ }: \{\det[E-H_{\mathrm{eff}}(\lambda)]\}^{[2]}= \prod_{n}[E- \epsilon_n^{_P}]-\lambda \sum_{n}h^{\!_P}_{nn}\prod_{n'\neq n}[E- \epsilon_{n'}^{_P}] +\lambda^2 \Big[ \\
	&\sum_{n\neq n'}\left( h^{\!_P}_{nn}h^{\!_P}_{n'n'}-|h^{\!_{P}}_{n n'}|^2\right) \prod_{n''\neq n,n'}[E- \epsilon_{n''}^{_P}]- \sum_{n,m} \frac{|h^{\!_{P\!Q}}_{nm}|^2}{\epsilon_n^{_P}-\epsilon_m^{_Q}}\prod_{n'\neq n}[E- \epsilon_{n'}^{_P}] \Big]=0; \\
	&\quad \cdots \cdots
\end{align*}
where $\{\cdots\}^{[K]}$ means that, to order of $ \lambda^K $, one only retains all the terms of $ \lambda^{k} $ with $ k \leq K $ \cite{C, Fried}.
Here we have not expanded product such as $\prod_{n=1}^{N}[E- \epsilon_n^{_P}]$, since if $N$ is large, the expansion result is very complicated.   

Obviously, such reconstruction of the effective secular equation is too cockamamie to adopt when the $P$-space has a large dimension $N$ (
Such trouble is also encountered for our treatment in \cite{Z}). This reconstruction is also  inefficient. Even to order of $ \lambda^2 $, one has to perform a second-order perturbation calculation for all the $ E_n^{_P} (\lambda)$, and some of these calculations actually are redundant since we have proved that terms such as those containing ``$\frac{1}{\epsilon_i^{_P}-\epsilon_n^{_P}}$'' must offset each other in the final effective secular equation.

\section{Improving procedure}
Here we show how to improve the usual implement of the reconstruction of an effective secular equation. Our strategy is to replace the characteristic polynomial by a ``relative'' one  
\begin{equation*}
	\det R (\lambda)\equiv\det\left[ \frac{E-H_{\mathrm{eff}}(\lambda)}{E-H_0^P}\right]= \det\left[1- \frac{V_{\mathrm{eff}}(\lambda)}{E-H_0^P}\right],
\end{equation*}
where the unperturbed $ P $-space Hamiltonian $H_0^P\equiv \sum_{n}\epsilon_{n}^{_P}|\psi^{_P}_n\rangle \langle \psi^{_P}_n|$ and $V_{\mathrm{eff}}(\lambda)\equiv H_{\mathrm{eff}}(\lambda)-H_0^P$. 

Obviously, such relative characteristic polynomial  $\det R (\lambda)=\det[E-H_{\mathrm{eff}}(\lambda)]/{\det[E-H_0^P]}$
has the same analytic property as $\det[E-H_{\mathrm{eff}}(\lambda)]$, as a function of $\lambda$, and hence they are completely equivalent for the the reconstruction of an effective secular equation. Actually, similar relative characteristic polynomials have been extensively used to perform a perturbation treatment of a Hamiltonian 
(though 
not for the purpose of establishing the effective secular equation), knowing as the Fredholm method or determinantal approach \cite{r1,r2,r3,r4,r5,r6}.

Next, we only need a specific form of $H_{\mathrm{eff}}(\lambda)$. According to our previous conclusion \cite{Z} that the analytic property $H_{\mathrm{eff}}(\lambda)$ has less effect on $\det[E-H_{\mathrm{eff}}(\lambda)]$, we in principle can choose any form of $H_{\mathrm{eff}}(\lambda)$ for the calculation of the relative characteristic polynomial; say, we can use the one resulted from the Rayleigh-Schr\"{o}dinger perturbation \cite{Bh1}, 
\begin{equation*}
	V_{\mathrm{eff}}(\lambda)=\lambda\sum_{n,n'} h^{\!_P}_{nn'}|\psi^{_P}_n\rangle \langle \psi^{_P}_{n'}|+\lambda^2 \sum_{m}\frac{h^{\!_{P\!Q}}_{nm}h^{\!_{P\!Q}}_{mn'}}{\epsilon_n^{_P}-\epsilon_m^{_Q}}|\psi^{_P}_n\rangle \langle \psi^{_P}_{n'}|+\cdots.
\end{equation*}
Then,
\begin{align*}
	&\text{to  $ \lambda^0$: } \{\det R (\lambda)\}^{[0]}=1; \\
	&\text{to  $ \lambda^1$ }: \{\det R (\lambda)\}^{[1]}= 1-\lambda \sum_{n}\frac{h^{\!_P}_{nn}}{E- \epsilon_{n}^{_P}};\\
	&\text{to  $ \lambda^2$ }: \{\det R (\lambda)\}^{[2]}= 1-\lambda \sum_{n}\frac{h^{\!_P}_{nn}}{E- \epsilon_{n}^{_P}}  \\
	& \qquad  \qquad +\lambda^2 \Big[ \sum_{n\neq n'}\frac{h^{\!_P}_{nn}h^{\!_P}_{n'n'}-|h^{\!_{P}}_{n n'}|^2}{[E- \epsilon_{n}^{_P}][E- \epsilon_{n'}^{_P}]} - \sum_{n,m} \frac{|h^{\!_{P\!Q}}_{nm}|^2}{\epsilon_n^{_P}-\epsilon_m^{_Q}} \frac{1}{E- \epsilon_{n'}^{_P}}\Big];\\
	&\quad \cdots \cdots
\end{align*}
Obviously, the corresponding effective secular equation $ \{\det R (\lambda)\}^{[K]} =0 $  ($K=1,2,\cdots$) is in a complete equivalence with the one $ \{\det[E-H_{\mathrm{eff}}(\lambda)]\}^{[K]} =0 $ shown above. 
However, the calculation with $ \{\det R (\lambda)\}^{[K]} =0 $ is much easier. Formally, $\{\det R (\lambda)\}^{[K]}$ has no trouble product such as $\prod_{n}[E- \epsilon_n^{_P}]$, greatly reducing the floating error, especially when $N$ is large. 

From the perspective of feasibility, one can perform a series expansion of $\det R (\lambda)$  using any equivalent expansion form of $H_{\mathrm{eff}}(\lambda)$, which is obviously much easier than the usual reconstruction of an effective secular equation from the
clumsy perturbation calculation of all the $ E_n^{_P} (\lambda)$. 
For a general $ \lambda $-series form of $V_{\mathrm{eff}}(\lambda)=\lambda V^{(1)}+\lambda^2 V^{(2)}+\cdots $, we can easily obtain
\begin{multline*}
	\det R (\lambda)= \det\left[1- \frac{V_{\mathrm{eff}}(\lambda)}{E-H_0^P}\right]=1-\sum_{n}\frac{\lambda V^{(1)}_{nn}}{E- \epsilon_{n}^{_P}} -\lambda^2\sum_{n}\frac{V^{(2)}_{nn}}{E- \epsilon_{n}^{_P}} \\
+\lambda^2 \sum_{n\neq n'}\frac{1}{[E- \epsilon_{n}^{_P}][E- \epsilon_{n'}^{_P}]}{	\begin{vmatrix}
		V^{(1)}_{nn}	& V^{(1)}_{nn'}\\
		V^{(1)}_{n'n}	& V^{(1)}_{n'n'}
\end{vmatrix} }+\cdots,
\end{multline*}
where $V^{(k)}_{nn'}=\langle \psi^{_P}_{n}|V^{(k)} |\psi^{_P}_{n'}\rangle $.

It is relatively easy to  obtain a  $\lambda$-expansion form of $H_{\mathrm{eff}}(\lambda)$ or $V_{\mathrm{eff}}(\lambda)$, comparing with that for the $ E_n^{_P} (\lambda)$. Actually, this has been widely studied and numerous methods have been developed to construct an effective Hamiltonian in a $\lambda$-series form \cite{Bh1, HS2,HS3,HS4,HS5}, which provide a good base for our improving procedure to reconstruct the effective secular equation. 
Generally speaking, different $\lambda$-expansion forms of  $H_{\mathrm{eff}}(\lambda)$ may hold different convergence radii.  It is generally very complicated to study the $\lambda$-expansion convergence radius for a  specific form of  $V_{\mathrm{eff}}(\lambda)$. However, different forms  of  $H_{\mathrm{eff}}(\lambda)$, which actually are similar matrices for each other, should result in the same characteristic polynomial $ \det[E-H_{\mathrm{eff}}(\lambda)]$ or $\det R (\lambda)$ via our procedure, yielding the same corresponding effective secular equation.

\section{Conclusion}
In summary, the usual procedure  for the reconstruction  of an effective secular equation can be greatly improved by introducing the relative characteristic polynomial.
Using our improving procedure, one can reconstruct an effective secular equation with any equivalent expansion form of $H_{\mathrm{eff}}(\lambda)$. The obtained effective-secular-equation  is equivalent to the usual one in Fried and Ezra's treatment, but is greatly simplified.  Our improving procedure shows great convenience and effectiveness when the $P$-space dimension is large.  Especially, to  overcome the  complex-energy-eigenvalue problem associated  with Fried and Ezra's treatment \cite{Fried,Z2}, one always needs higher-order $\lambda$-expansion form of the effective secular equation, but the usual  reconstruction procedure would be too cockamamie to implement; however, our improving procedure may show its usefulness, due to its simplicity.

\end{document}